\documentclass[5p,twocolumn]{elsarticle}

\usepackage{lineno,hyperref}
\usepackage{txfonts}
\usepackage{multirow}
\usepackage{color}
\usepackage{dcolumn}
\usepackage{ulem} 
\modulolinenumbers[5]

\journal{the Journal of Alloys and Compounds}

\begin{document}

\begin{frontmatter}

\title{Synthesis and study of transport and magnetic properties of magnesium cage compounds
$R$Ni$_{2}$Mg$_{20}$ ($R$ = Pr and Nd)}


\author[AdSE]{Yuka Kusanose}
\author[AdSE]{Takahiro Onimaru\corref{mycorrespondingauthor}}
\cortext[mycorrespondingauthor]{Corresponding author}
\ead{onimaru@hiroshima-u.ac.jp}
\author[AdSE]{Yu Yamane\corref{presentaddress}}
\cortext[presentaddress]{Department of Material Science, Graduate School of Science, University of Hyogo, Kamigori, Hyogo 678-1297, Japan}
\author[NBARD]{Kazunori Umeo}
\author[AdSE]{Toshiro Takabatake}

\address[AdSE]{Department of Quantum Matter, Graduate School of Advanced Science and Engineering, Hiroshima University, 
Higashi-Hiroshima 739--8530, Japan}
\address[NBARD]{Department of Low Temperature Experiment, Integrated Experimental Support / Research Division, N-BARD, Hiroshima University, Higashi-Hiroshima 739--8526, Japan}

\begin{abstract}
We have synthesized magnesium cage compounds $R$Ni$_{2}$Mg$_{20}$ ($R$ = Pr and Nd). 
We report the measurements of electrical resistivity $\rho$, magnetic susceptibility $\chi$, isothermal magnetization $M$, and specific heat $C$.
Polycrystalline samples with $R$ $=$ Pr and Nd were obtained by annealing, while the counterparts for $R$ $=$ La, Ce, and Y were found to be absent.
The $\rho(T)$ data for $R$ $=$ Pr and Nd monotonically decrease on cooling from 300 K to 40 K and exhibit shoulders at around 13 K and 15 K, respectively.
The Curie--Weiss behaviors of the $\chi(T)$ data indicate the trivalent states of both the Pr and Nd ions. 
For $R$ $=$ Pr, the maximum in $C(T)$  at around 7 K is reproduced by a doublet-triplet two-level model with an energy gap of 14 K. 
The peak in $C(T)$ at 0.7 K is attributed to a short-range order of quadrupolar degrees of freedom in the non-Kramers ground doublet. 
On the other hand, for $R$ = Nd, a maximum in $C(T)$ at around 9 K is explained by thermal excitation from a ground state doublet to an excited quartet separated by 23 K.
Upon further cooling, $C(T)$ shows a lambda-shaped peak at 1.5 K.
By applying magnetic fields up to 2 T, the peak becomes broad and shifts to higher temperatures, which is a characteristic of a ferromagnetic order.
\end{abstract}


\end{frontmatter}


\section{Introduction}

Recently, there has been considerable interest in a class of cage-structured compounds $R$$T_2$$X_{20}$ where $R$ is a rare-earth element, $T$ is a transition metal, and $X$ $=$ Al, Zn, and Cd.
They crystallize in the cubic CeCr$_{2}$Al$_{20}$-type structure with the space group of $Fd\bar{3}m$ ($O_{h}^{7}$, \#227), where the $R$ and $T$ atoms at the 8$a$ and 16$d$ sites form the diamond and pyrochlore sublattices, respectively \cite{Nasch97}.
YbCo$_{2}$Zn$_{20}$ with 4$f^{13}$ configuration exhibits a heavy fermion state with an extremely large electronic specific heat coefficient of $\gamma$ $=$ 8 J$/$K$^2$ mol \cite{Torikachvili07,Saiga08,Honda14}.
Coexistence of quadrupole order and superconductivity observed in the non-Kramers doublet systems Pr$T$$_2$Zn$_{20}$ ($T$ $=$ Ir and Rh) \cite{Pr1-2-20,Onimaru10,Onimaru11,Onimaru12,Umeo20} and Pr$T$$_2$Al$_{20}$ ($T$ $=$ Ti and V) \cite{Sakai11,Sakai12,Matsubayashi12,Tsujimoto14} suggests that the superconducting pair is mediated by quadrupole fluctuations. 
Above the quadrupole ordering temperatures, non-Fermi liquid (NFL) behaviors were observed in Pr$T$$_2$Zn$_{20}$ ($T$ $=$ Ir and Rh) \cite{Onimaru16,Yamada19,Yoshida17} and PrV$_{2}$Al$_{20}$ \cite{Fu20}.
Furthermore, the NFL behaviors in the diluted Pr system Y(Pr)Ir$_2$Zn$_{20}$ \cite{Yamane18a,Yanagisawa19} indicate manifestation of the quadrupole Kondo effect \cite{Cox98,Tsuruta15}.
On the other hand, in isostructural Nd-based compounds Nd$T_2$$X_{20}$ with the 4$f^3$ configuration, the crystalline electric field (CEF) ground states are mostly the magnetic $\Gamma_6$ doublets \cite{Isikawa13,Wakiya15,Wakiya17,Yamane17,Yamamoto19,Namiki16}. 
Nd$T_2$$X_{20}$ exhibit either ferromagnetic (FM) or antiferromagnetic (AFM) transitions. 
Thereby, the competing FM and AFM interactions between the Nd ions forming the diamond sublattice may induce magnetic fluctuations to hinder the magnetic long-range order. 
In addition, a theoretical model predicts that Nd$T_2$$X_{20}$ compounds with the $\Gamma_6$ doublet ground states have potential exhibiting the two-channel Kondo effect \cite{Hotta17}.

\begin{figure}
	\centering
	\includegraphics[scale=0.2]{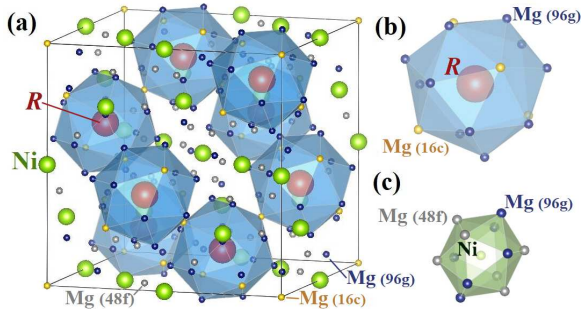}
	\caption{(Color online) (a) Crystal structure of $R$Ni$_2$Mg$_{20}$ ($R$ $=$ Pr and Nd) crystalizing in the cubic CeCr$_2$Al$_{20}$-type with the space group of $Fd\bar{3}m$ with $Z$ $=$ 8 \cite{Nasch97}. 
(b) An $R$ ion is surrounded by a Frank--Kasper polyhedra consisting of four Mg atoms at the 16$c$ site and twelve Mg atoms at the 96$g$ site.
The point group of the $R$ site is the cubic $T_d$.
(c) Icosahedron formed by Mg atoms at the 48$f$ site and Mg atoms at the 96$g$ sites which surround the Ni atom.
}
	\label{fig1}
\end{figure}

An isostoichiometric magnesium compound NdNi$_{2}$Mg$_{20}$ was synthesized as a potential material with efficient hydriding/dehydriding properties \cite{Luo15,Li17}. 
The crystal structure was refined to be a tetragonal one with the space group of $I4_{1}/amd$ ($D_{4h}^{19}$, \#141) with the lattice parameters of $a$ $=$ 11.2743 {\AA}, $c$ $=$ 15.9170 {\AA}, and $Z$ $=$ 4 \cite{Luo15}.
This tetragonal structure could be transformed from the cubic CeCr$_{2}$Al$_{20}$-type one with the space group of $Fd\bar{3}m$ and $Z$ $=$ 8 (Fig. \ref{fig1}) as follows:
The 48$f$ site of the Mg atoms in the cubic structure is split into the 16$g$ and 8$e$ sites of the tetragonal one with $Z$ $=$ 4, by which the three-fold symmetry in the cubic structure is broken and the unit cell volume is decreased to a half.

In the present work, we synthesized polycrystalline samples of $R$Ni$_2$Mg$_{20}$ of $R$ $=$ Pr and Nd and measured the X-ray diffraction (XRD) patterns to analyze the crystal structures.
The trivalent states of the Pr and Nd ions have been identified by the effective magnetic moments evaluated from the magnetic susceptibility.
Assuming that the point group at the Pr site was cubic, the CEF ground state could be a non-Kramers doublet. 
However, no phase transition was observed at low temperatures down to 0.1 K.
For $R$ $=$ Nd, the Nd ions with a Kramers doublet ground state exhibit a ferromagnetic transition.

\begin{figure}
 \begin{minipage}{0.49\hsize}
  \begin{center}
   \includegraphics[scale=0.15]{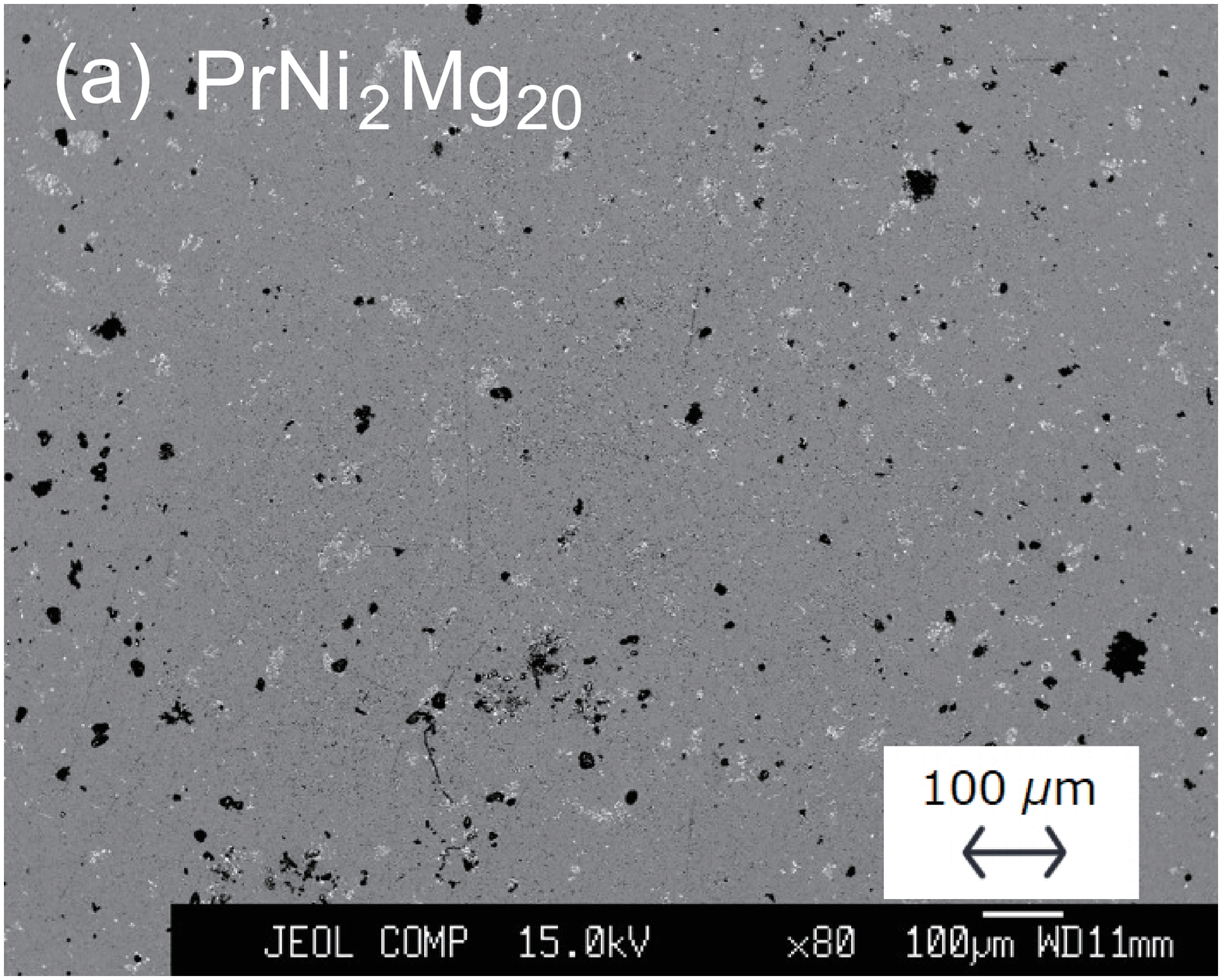}
  \end{center}
 \end{minipage}
 \begin{minipage}{0.49\hsize}
  \begin{center}
   \includegraphics[scale=0.15]{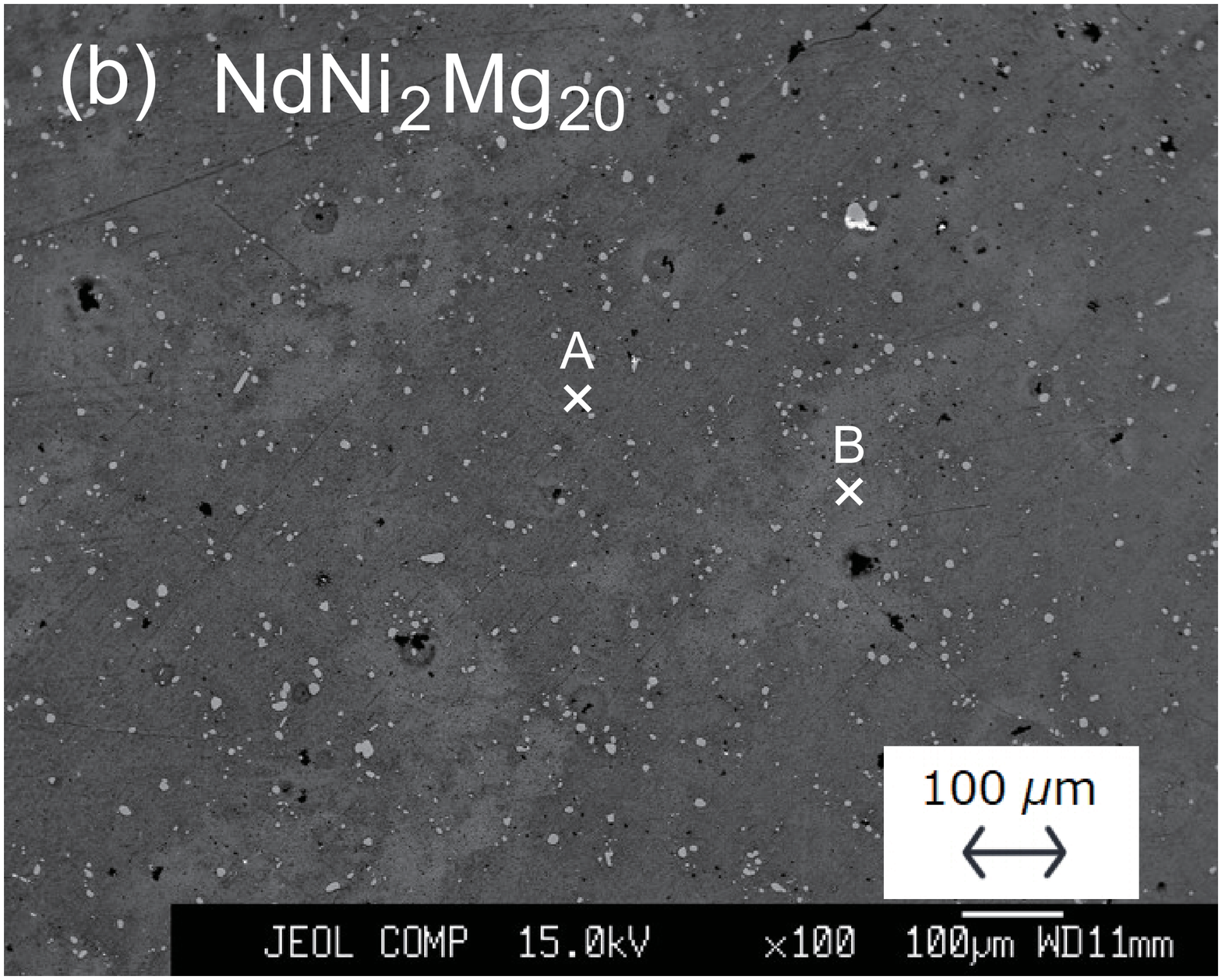}
  \end{center}
 \end{minipage}
 \caption{Backscattered electron images of the samples of $R$Ni$_2$Mg$_{20}$ for (a) $R$ $=$ Pr and (b) Nd.
(a) The atomic composition determined by EPMA is PrNi$_{1.82(2)}$Mg$_{20.3(1)}$ for the primary phase. Impurity phases are colored white, and the black areas are holes.
(b) The primary phase colored dark gray (labeled with ``A") is NdNi$_{1.9(2)}$Mg$_{19.7(10)}$. Another phase colored light gray (labeled with ``B") is NdNi$_{1.9(1)}$Mg$_{16.0(13)}$, where some of the Mg atoms may be deficient.
An impurity phase colored white is a binary Ni--Mg one.
 }
 \label{fig_SEM}
\end{figure}

\section{Syntheses and characterization of samples}

Polycrystalline samples of $R$Ni$_{2}$Mg$_{20}$ ($R$ $=$ Pr and Nd) were prepared by using high-purity rare-earth ingots of 99.9\% Pr (Johnson Matthey) and 99.9\% Nd (Ames laboratory), Ni rods of 99.995\% (Johnson Matthey), and Mg shots of 99.99\% (Rare Metallic).
First, binary alloys of $R$Ni$_2$ were prepared by the arc-melting in an argon atmosphere. Next, the $R$Ni$_2$ alloys and Mg shots with the stoichiometric ratio were placed in an alumina crucible which was sealed in a quartz tube with an argon atmosphere at pressure of about 1$/$3 atm. 
The mixtures were heated with a high-frequency induction furnace, and then quenched in water. 
The reactant was wrapped in Ta foil and resealed in a quartz tube for annealing in an electric furnace.
Polycrystalline samples of $R$Ni$_{2}$Mg$_{20}$ for $R$ $=$ Pr and Nd were obtained by annealing at the temperatures between 350 and 450$^{\circ}$C for 7 days.
Crystal structures of the annealed samples were analyzed by using a program RIETAN-FP to fit the XRD patterns \cite{Izumi07}.
For the main phases, the atomic compositions were determined by averaging over 10 different spots for each sample by means of the electron-probe micro-analysis (EPMA) with a JEOL JXA-8200 analyzer. 
On the other hand, syntheses of samples for $R$ $=$ Y, La, and Ce were failed by the same manner with using rare-earth ingots of 99.9\% Y, La and Ce (Johnson Matthey). 
Details are described in the following paragraphs.

\begin{figure}
	\centering
	\includegraphics[scale=0.32]{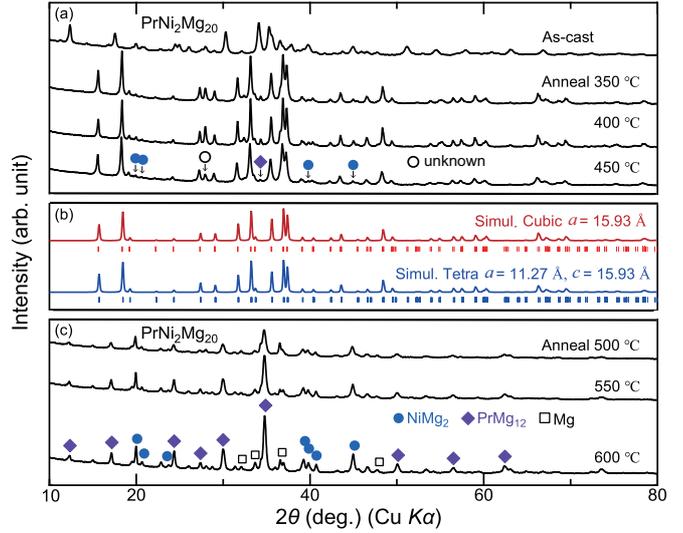}
	\caption{(Color online)
(a) (c) Powder X-ray diffraction patterns of the as-cast sample of PrNi$_2$Mg$_{20}$ and those annealed at temperatures between 350 and 600$^{\circ}$C for 7 days.
(b) The upper (red) is the simulated pattern with the cubic CeCr$_2$Al$_{20}$-type structure.
The lower is the simulated pattern (blue) with the tetragonal structure with the space group of $I4_{1}/amd$ as proposed in ref. \cite{Luo15}, see text for details.
Most of the diffraction peaks of the samples annealed at 350, 400, and 450$^{\circ}$C shown in (a) can be indexed by the cubic CeCr$_2$Al$_{20}$-type structure, 
while additional weak peaks due to impurity phases are shown with the arrows.
The cubic phase disappears in the samples annealed at $T$ $\ge$ 500$^{\circ}$C, decomposing into NiMg$_2$, PrMg$_{12}$, and Mg.
  }
	\label{fig3}
\end{figure}

Figure \ref{fig_SEM} shows backscattered electron images of the polycrystalline $R$Ni$_{2}$Mg$_{20}$ samples of (a) $R$ $=$ Pr and (b) Nd.
For $R$ $=$ Pr, the EPMA identified the main phase as PrNi$_2$Mg$_{20}$.  
Assuming full occupancy of the Pr site, the composition was obtained as PrNi$_{1.82(2)}$Mg$_{20.3(1)}$.
On the other hand, for $R$ $=$ Nd (b), besides the main phase colored dark gray with the label ``A", another phase colored light gray with the label ``B" was detected in moderately wide area.
Assuming full occupancy of the Nd site, the composition of the phase A is estimated to be NdNi$_{1.9(2)}$Mg$_{19.7(10)}$, while that of the phase B is NdNi$_{1.9(1)}$Mg$_{16.0(13)}$.
The large off-stoichiometry in the phase B implies inhomogeneous distribution of the atomic compositions.
Since the XRD pattern of the powdered sample indicates the cubic crystal structure as described below, some of the Mg sites in the phase B may be deficient.
An impurity phase colored white is a binary Ni--Mg phase.


Figures \ref{fig3}(a) and \ref{fig3}(c) show the powder X-ray diffraction patterns of PrNi$_2$Mg$_{20}$ samples of as-cast form and annealed forms at 350--600$^{\circ}$C for 7 days.
Here, it is noted that, in the previous report on NdNi$_{2}$Mg$_{20}$, the crystal structure was proposed as the tetragonal one with the space group of $I4_{1}/amd$ \cite{Luo15}.
In Fig. \ref{fig3}(b), we compare the simulated XRD patterns for the cubic CeCr$_2$Al$_{20}$-type and the tetragonal crystal structures.
The pattern of the as-cast sample in Fig. \ref{fig3}(a) is much different from the simulated patterns,
whereas those of the samples annealed at 350, 400, and 450$^{\circ}$C are indexed with both the cubic and tetragonal structures.
Since no difference between the two simulated patterns is recognized, we adopt the cubic crystal structure hereafter. 
By the Rietveld refinement, the cubic lattice parameter $a$ of the samples annealed between 350 and 450$^{\circ}$C was estimated to be 15.944(1) {\AA}.
This value is much larger than 15.575(1) and 15.699(1) {\AA} of the isostructural Pr$T_{2}$Cd$_{20}$ of $T$ $=$ Ni and Pd, respectively \cite{Burnett14}.
Thereby, negative chemical pressure due to the larger lattice parameter is expected to weaken the CEF effect on the Pr ion in PrNi$_2$Mg$_{20}$.

\begin{figure}
	\centering
	\includegraphics[scale=0.32]{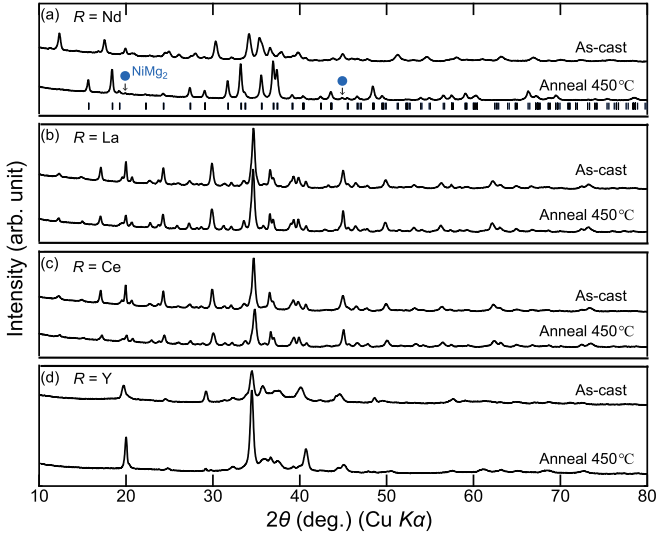}
	\caption{(Color online)
Powder X-ray diffraction patterns of samples of $R$Ni$_2$Mg$_{20}$ for (a) $R$ $=$ Nd, (b) La, (c) Ce, and (d) Y for as-cast forms and those annealed at 450$^{\circ}$C for 7 days.
Among them, only the annealed sample for $R$ $=$ Nd is identified as the expected cubic phase, whereas weak peaks shown with the arrows due to the impurity of NiMg$_{2}$ were observed.
}
	\label{fig4}
\end{figure}

As the annealing temperature is increased above 500$^{\circ}$C, the diffraction pattern changes to those composed of NiMg$_2$, PrMg$_{12}$, and Mg as shown in Fig. \ref{fig3}(c) with the (blue) closed circles, (purple) diamonds, and opened squares, respectively. 
This temperature dependence indicates that the PrNi$_2$Mg$_{20}$ phase decomposes for $T$ $\ge$ 500$^{\circ}$C.
Looking back at the patterns of the samples annealed at $T$ $\le$ 450$^{\circ}$C shown in Fig. \ref{fig3}(a), small peaks due to the impurity phases of NiMg$_2$ and PrMg$_{12}$ are found. 
Another peak is found at 28$^{\circ}$ as shown with the opened circle, which is ascribed to an unknown reactant.
Since the fractions of the impurity phases are tiny, the low temperature physical properties may not be affected by the impurity phases.

Figure \ref{fig4}(a) shows the powder XRD patterns of the $R$ = Nd samples of as-cast form and annealed one at 450$^{\circ}$C for 7 days.
The latter pattern is well indexed as the cubic phase of NdNi$_{2}$Mg$_{20}$,
while small impurity peaks of NiMg$_{2}$ were detected as indicated by the arrows.
The cubic lattice parameter was estimated to be 15.943(2) \AA, which is larger than 15.9170 {\AA} reported previously \cite{Luo15}.
Figures \ref{fig4}(b)--\ref{fig4}(d) show the powder XRD patterns of $R$Ni$_{2}$Mg$_{20}$ samples for $R$ $=$ La, Ce, and Y of the as-cast form and annealed form at 450$^{\circ}$C for 7 days.
The annealing does not affect the XRD pattern. Neither of the patterns contains the peaks expected for the cubic $R$Ni$_{2}$Mg$_{20}$.

\begin{figure}[h]
\centering
\includegraphics[scale=0.32]{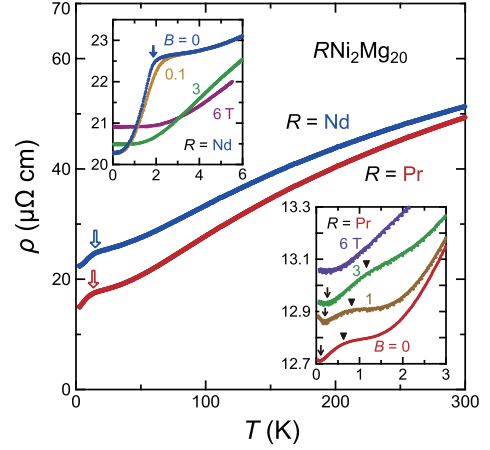}
\caption{(Color online) Temperature dependence of the electrical resistivity $\rho(T)$ of PrNi$_2$Mg$_{20}$ (red) and NdNi$_2$Mg$_{20}$ (blue). 
The arrows mark shoulders at around 13 K and 15 K.
The upper inset shows the $\rho(T)$ data for $R$ $=$ Nd in magnetic fields of $B$ $=$ 0, 0.1, 3, and 6 T. 
For $B$ $=$ 0, a sharp bend at 1.9 K (blue arrow) indicates a phase transition.
The lower inset shows the $\rho(T)$ data of $R$ $=$ Pr in magnetic fields of $B$ $=$ 0, 1, 3, and 6 T.
The data in the magnetic fields are vertically offset for clarity.
The triangle at 0.7 K for $B$ $=$ 0 marks a broad shoulder, which becomes broader and shifts to higher temperatures with increasing $B$ to 3 T.
A minimum at around 0.1 K in $B$ $=$ 0 marked by the arrow becomes more distinct at $B$ $=$ 1 T but disappears at 6 T.
}
\label{f_rho_01}
\end{figure}

\section{Physical property measurements}

\subsection{Experimental}

Electrical resistance was measured by a standard four-probe AC method with a laboratory-built system installed in a Gifford-McMahon-type refrigerator in the temperature range of 3 $<$ $T$ $<$ 300 K and with a commercial Cambridge Magnetic Refrigerator mFridge for 0.1 $<$ $T$ $<$ 4 K. 
The magnetization was measured from 1.8 to 300 K in magnetic fields $B$ $\le$ 5 T by using a commercial superconducting quantum interference device (SQUID) magnetometer MPMS (Quantum Design). 
Specific heat was measured by the thermal relaxation method with a Quantum Design physical property measurement system (PPMS) for 0.4 $<$ $T$ $<$ 20 K and with the mFridge for 0.1 $<$ $T$ $<$ 0.6 K.

\subsection{Electrical resistivity}

Temperature variations of the electrical resistivity $\rho(T)$ of the samples of $R$Ni$_2$Mg$_{20}$ ($R$ $=$ Pr and Nd) annealed at 400$^{\circ}$C for 7 days are shown in Fig. \ref{f_rho_01}.
The values of residual resistivity ratio (RRR) evaluated by $\rho$(300 K)$/$$\rho$(2.8 K) are 3.9 and 2.5 for $R$ $=$ Pr and Nd, respectively.
On cooling from 300 to 40 K, $\rho(T)$ monotonically decreases.
There are shoulders at around 13 K and 15 K, respectively, for $R$ $=$ Pr and Nd as shown with the arrows.
The shoulders probably arise from scattering of conduction electrons by the 4$f$ electrons in the thermally excited CEF levels of the Pr and Nd ions \cite{Abou75}, because the temperatures are comparable to the values of excitation energy from the CEF ground states to the excited states, as will be discussed later.

For $R$ $=$ Pr, $\rho(T)$ exhibits a broad shoulder at 0.7 K and a minimum at 0.1 K as shown with the triangle and arrow, respectively, in the lower inset of Fig. \ref{f_rho_01}.
No clear transition was observed down to the lowest temperature of 0.04 K.
The data in the magnetic fields are vertically offset for clarity.
The shoulder at 0.7 K is robust against magnetic fields $B$ up to 1 T.
The shoulder becomes broad and shifts to 1.2 K at $B$ $=$ 3 T.
On the other hand, the minimum at 0.1 K becomes more distinct at $B$ $=$ 1 T and disappears at $B$ $=$ 6 T.
This peculiar behavior is likely relevant to the increase of the specific heat divided by temperature $C/T$ on cooling below 0.2 K as will be discussed later.

For $R$ $=$ Nd, as shown with the arrow in the upper inset, $\rho(T)$ sharply bends at 1.9 K.
It is ascribed to a phase transition since the specific heat $C(T)$ shows a lambda-type anomaly in the vicinity of the temperature as will be described later.
The bend becomes broad by applying magnetic field of $B$ $=$ 0.1 T and disappears at 6 T, suggesting the ferromagnetic nature of the transition.

\begin{figure}
\centering
\includegraphics[scale=0.45]{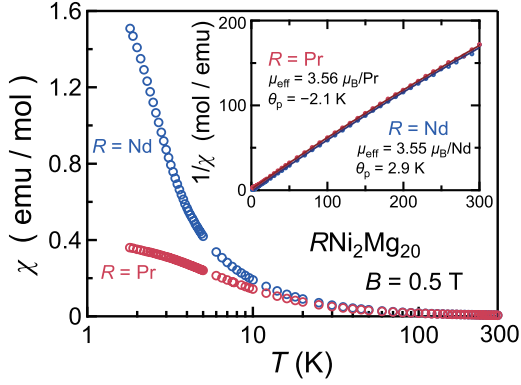}
\vspace{2mm}
\caption{(Color online) Temperature dependence of the magnetic susceptibility $\chi(T)$ of $R$Ni$_{2}$Mg$_{20}$ for $R$ = Pr and Nd in a magnetic field of $B$ $=$ 0.5 T.
The inset shows the inverse $\chi^{-1}(T)$. 
The solid lines are fits with a modified Curie--Weiss form, see text in detail.
}
\label{f_chi_01}
\end{figure}

\subsection{Magnetic susceptibility and magnetization}

We used samples of $R$Ni$_2$Mg$_{20}$ ($R$ $=$ Pr and Nd) annealed at 450$^{\circ}$C for 7 days for the measurements of the temperature variations of the magnetic susceptibility $\chi(T)$, isothermal magnetization $M(B)$, and the specific heat $C(T)$.

Figure \ref{f_chi_01} shows $\chi(T)$ in a magnetic field of $B$ $=$ 0.5 T and the inset shows the inverse $\chi^{-1}(T)$.
As shown with the solid lines, $\chi^{-1}(T)$ between 50 and 300 K follows a modified Curie--Weiss form expressed by $\chi(T)$ $=$ $C$$/$($T$ $-$ $\theta_{\rm p}$) $+$ $\chi_{0}$, where $C$ is a Curie constant, $\theta_{\rm p}$ a paramagnetic Curie temperature, and $\chi_{0}$ temperature independent susceptibility.
The fit to the $\chi^{-1}(T)$ data of $R$ = Pr gives the values of $\theta_{\rm p}$ $=$ $-$2.1 K and $\chi_{0}$ $=$ 6.2 $\times$10$^{-4}$ emu/mol.
The negative $\theta_{\rm p}$ indicates the antiferromagnetic intersite interaction between the Pr moments.
The effective magnetic moment was estimated to be $\mu_{\rm eff}$ $=$ 3.56 $\mu_{\rm B}$/f.u., which is close to 3.58 $\mu_{\rm B}$ for a free trivalent Pr ion.
On the other hand, for $R$ = Nd, the Curie--Weiss fit gives the values of $\theta_{\rm p}$ $=$ $+$2.9 K and $\chi_{0}$ $=$ 6.4 $\times$10$^{-4}$ emu/mol.
The positive value of $\theta_{\rm p}$ indicates the ferromagnetic intersite interaction between the Nd moments.
The obtained value of $\mu_{\rm eff}$ $=$ 3.55 $\mu_{\rm B}$/f.u. is close to 3.62 $\mu_{\rm B}$ for a free trivalent Nd ion.

As shown in the main panel, the increase in $\chi(T)$ for $R$ $=$ Pr is moderately suppressed on cooling below 5 K, indicating van-Vleck paramagnetic behavior expected for a nonmagnetic CEF ground state.
In contrast, $\chi(T)$ for $R$ $=$ Nd exhibits divergent behavior on cooling below 5 K, which arises from the magnetic degrees of freedom in a Kramers ground multiplet.

\begin{figure}
 \begin{minipage}{0.49\hsize}
  \begin{center}
   \includegraphics[scale=0.28]{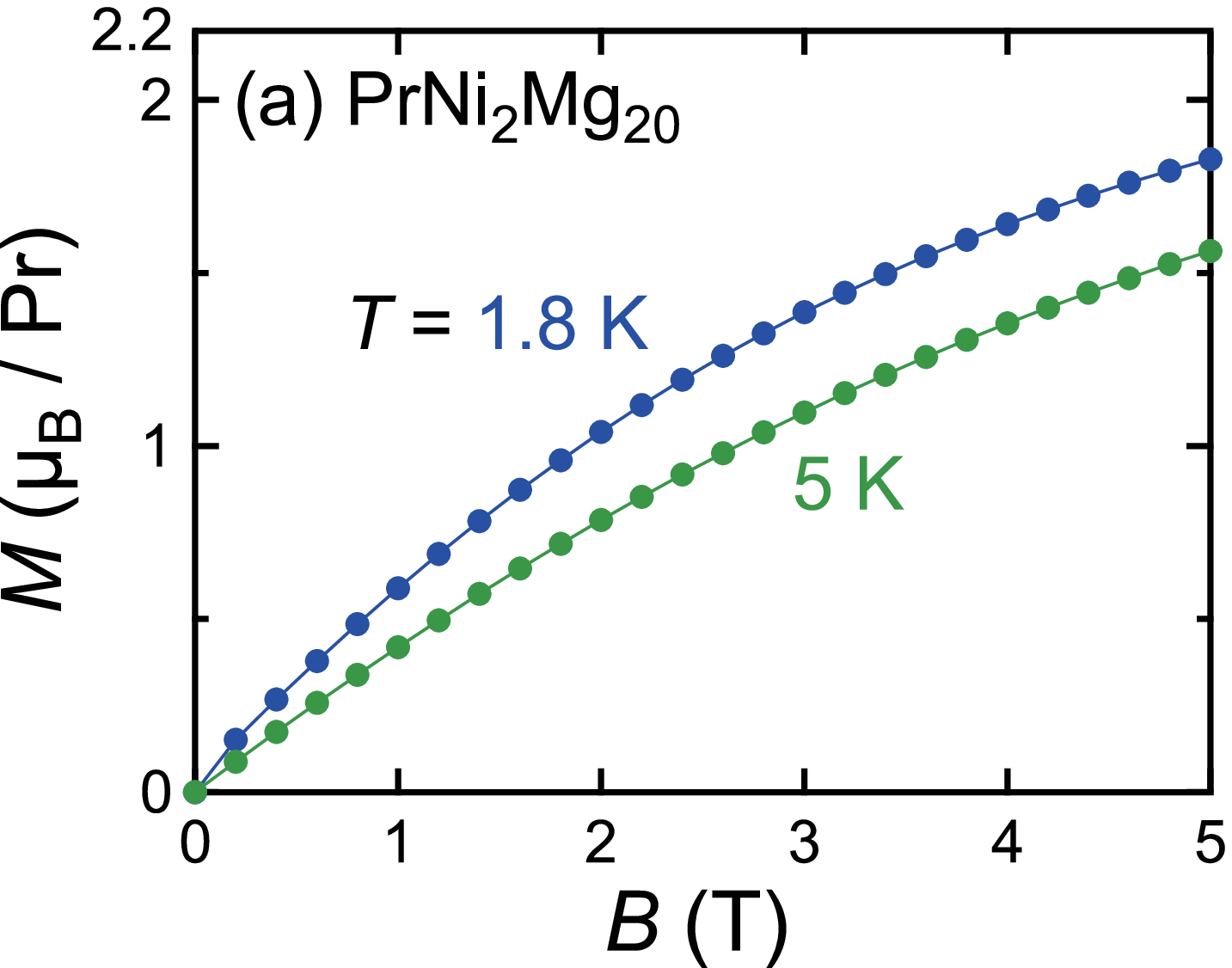}
  \end{center}
 \end{minipage}
 \begin{minipage}{0.49\hsize}
  \begin{center}
   \includegraphics[scale=0.28]{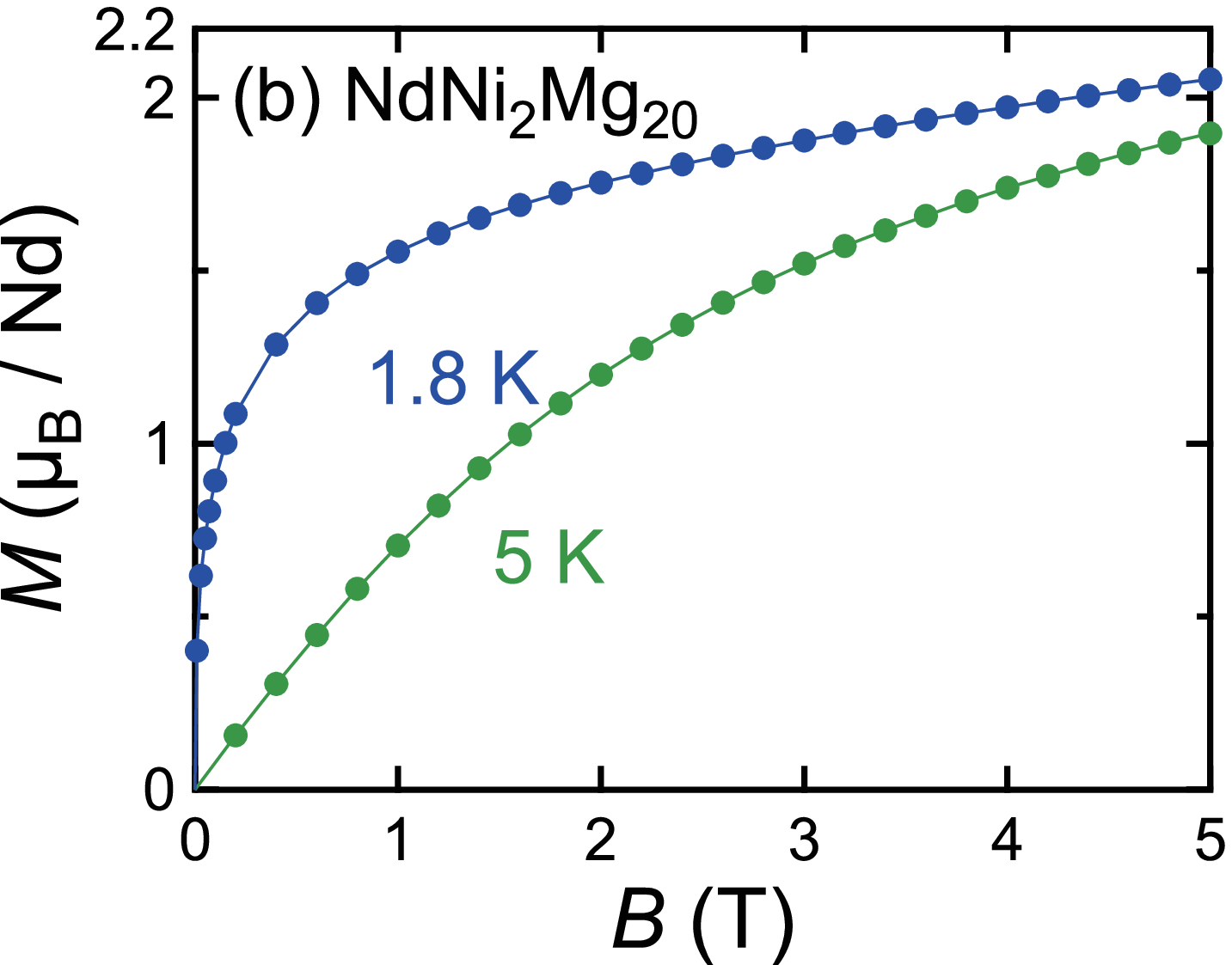}
  \end{center}
 \end{minipage}
\caption{
(Color online) Isothermal magnetization $M(B)$ of $R$Ni$_{2}$Mg$_{20}$ for (a) $R$ = Pr and (b) Nd at $T$ $=$ 1.8 and 5 K. 
}
\label{fig_M}
\end{figure}

Figures \ref{fig_M}(a) and \ref{fig_M}(b) display $M(B)$ of $R$Ni$_{2}$Mg$_{20}$ for $R$ $=$ Pr and Nd, respectively, at $T$ $=$ 1.8 and 5 K.
The $M(B)$ data for $R$ $=$ Pr monotonically increase with increasing $B$, suggesting a paramagnetic state.
This is consistent with the van-Vleck paramagnetic behavior of $\chi(T)$ shown in Fig. \ref{f_chi_01}.
In contrast, as shown in Fig. \ref{fig_M}(b), $M(B)$ for $R$ $=$ Nd at 1.8 K first rapidly increases from the origin, and then gradually increases for $B$ $\ge$ 1 T.
Since the temperature of 1.8 K is just below the transition temperature of 1.9 K,
the rapid increase of $M(B)$ from the origin is ascribed to the ferromagnetic correlation between the Nd moments expected from the positive value of $\theta_{\rm p}$.
At $T$ $=$ 5 K, $M(B)$ monotonically increases with increasing $B$, indicating the paramagnetic state above the transition temperature.

\begin{figure}
 \begin{minipage}{0.49\hsize}
  \begin{center}
   \includegraphics[scale=0.32]{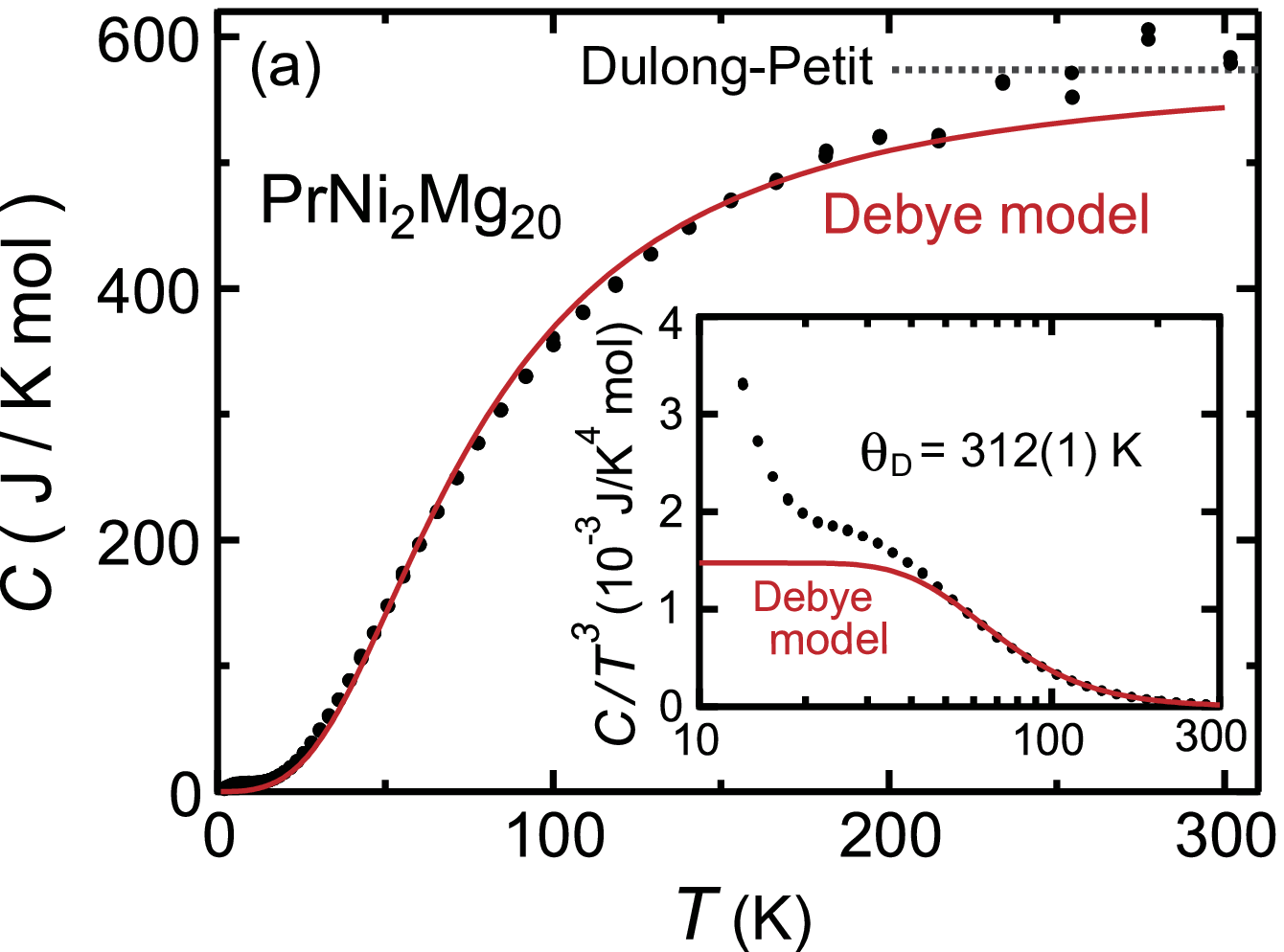}
  \end{center}
 \end{minipage}
 \begin{minipage}{0.49\hsize}
  \begin{center}
   \includegraphics[scale=0.32]{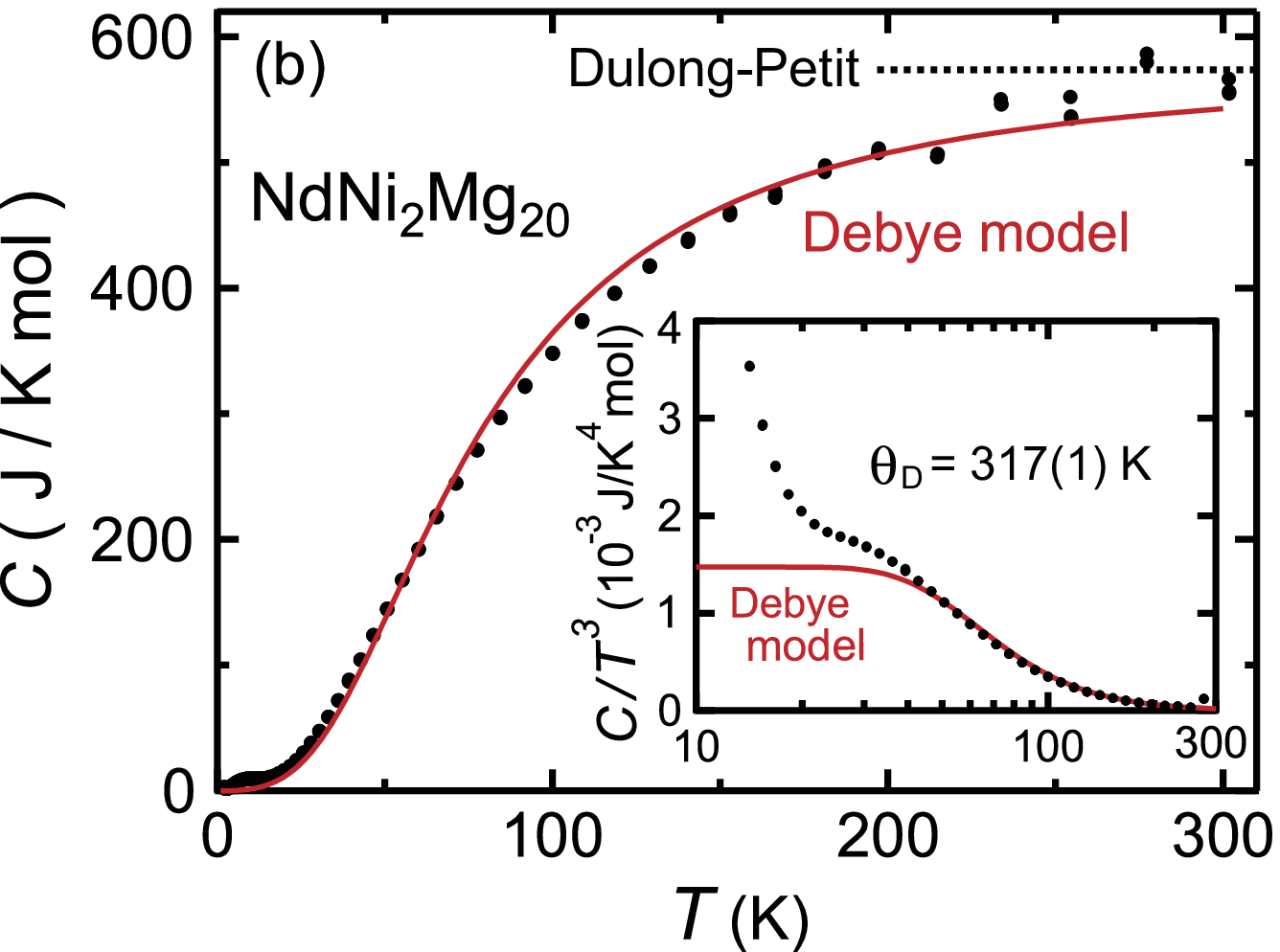}
  \end{center}
 \end{minipage}
\caption{
(Color online) Specific heat $C(T)$ of $R$Ni$_{2}$Mg$_{20}$ for (a) $R$ $=$ Pr and (b) Nd between 2 and 300 K. 
The insets show the $C/T^3$ vs $T$ plots. The (red) solid lines indicate simulations by adopting the Debye model with the Debye temperatures of ${\theta}_{\rm D}$ $=$ 312(1) and 317(1) K for $R$ $=$ Pr and Nd, respectively. 
}
\label{fig_HC}
\end{figure}

\subsection{Specific heat and magnetic entropy}

Specific heat $C(T)$ data of $R$Ni$_{2}$Mg$_{20}$ for $R$ $=$ Pr and Nd between 2 and 300 K are shown in Figs. \ref{fig_HC}(a) and \ref{fig_HC}(b), respectively.
Both the data increase with elevating temperatures and approach the Dulong-Petit value of 3$nR$ $=$ 573.4 J$/$K mol, where $n$ $=$ 23 is the number of atoms per formula unit and $R$ the gas constant.
The (red) solid curves are the fits to the $C(T)$ data for 50 $<$ $T$ $<$ 300 K by adopting the Debye model for acoustic phonon modes \cite{Kittel} with the Debye temperatures of ${\theta}_{\rm D}$ $=$ 312(1) K for $R$ $=$ Pr and 317(1) K for $R$ $=$ Nd.
The values of ${\theta}_{\rm D}$ are higher than that of ${\theta}_{\rm D}$ $=$ 259 K for an isostructural LaRu$_{2}$Zn$_{20}$ \cite{Wakiya16}, which is reasonable because the molecular weights of $R$Ni$_2$Mg$_{20}$ for $R$ $=$ Pr and Nd are much lighter than that of LaRu$_{2}$Zn$_{20}$.
The insets display the $C/T^3$ plots with the logarithmical scale. 
The data agree well with the simulations of the Debye model for 50 $<$ $T$ $<$ 300 K, whereas they deviate from the simulated curves for $T$ $<$ 50 K, which is ascribed to the magnetic contributions of the 4$f$ electrons. 
While optical phonon modes were observed at around 15 K in the $C/T^3$ plot for the nonmagnetic LaRu$_{2}$Zn$_{20}$ \cite{Wakiya16}, the large contribution of the 4$f$ electrons apparently hinders the optical phonon modes in the present Pr and Nd substances.

\begin{figure}
\centering
\includegraphics[scale=0.34]{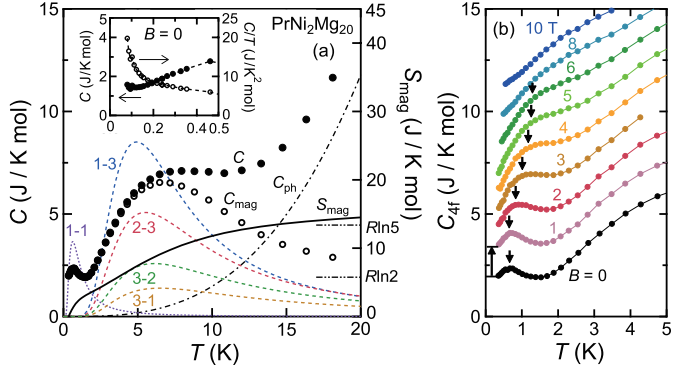}
\caption{(Color online) (a) Temperature dependence of the specific heat $C$, magnetic specific heat $C_{\rm mag}$, and magnetic entropy $S_{\rm mag}$ of PrNi$_2$Mg$_{20}$. 
$C_{\rm mag}$ shown with the opened circles was estimated by subtracting the phonon contribution $C_{\rm ph}$ (dot-dashed line) from the $C(T)$ data.
The dashed curves labeled ``1-3" (blue)", ``2-3" (red), ``3-2" (green), and ``3-1" (yellow) are calculations with two-level models of singlet-triplet, doublet-triplet, triplet-doublet, and triplet-singlet separated by 14 K, respectively. 
The (purple) dotted curve (1-1) displays a calculation of a singlet-singlet two-level model with an energy gap of 1.5 K.
The inset shows the $C(T)$ (left-hand scale) and $C/T$ (right-hand scale) data from 0.08 K to 0.5 K.
(b) The 4$f$ contribution to the specific heat, $C_{4f}$, in magnetic fields up to $B$ $=$ 10 T.
To evaluate $C_{4f}$, the phonon and nuclear contributions were subtracted from the measured $C(T)$ data.
The data are vertically offset for clarity.
With increasing $B$, a broad peak at 0.7 K in $B$ $=$ 0 becomes broad and reaches 1.3 K at 6 T. 
}
\label{fig_HC_Pr}
\end{figure}

The temperature variation of $C$ of PrNi$_2$Mg$_{20}$ for $T$ $<$ 20 K is shown in Fig. \ref{fig_HC_Pr}(a).
Because the nonmagnetic counterparts $R$Ni$_2$Mg$_{20}$ for $R$ $=$ La and Y could not be synthesized, 
the magnetic specific heat $C_{\rm mag}$ was estimated by subtracting the phonon contribution $C_{\rm ph}$ from the total $C(T)$ data, where $C_{\rm ph}$ was
simulated by adopting the Debye model with ${\theta}_{\rm D}$ $=$ 312 K as described above.
As shown with the open circles in Fig. \ref{fig_HC_Pr}(a), $C_{\rm mag}$ exhibits a maximum at around 7 K, which probably results from the Schottky specific heat due to thermal excitation from the CEF ground state to the low-lying levels as described below.

The Schottky specific heat of a two-level model is expressed by an equation as follows. 
\begin{equation}
	C = \frac{n m \Delta^2}{k_{\rm B} T^2}  \frac{\mathrm{e}^{- {\Delta} / {k_{\rm B} T}}}{(n + m \mathrm{e}^{- {\Delta} / {k_{\rm B} T}})^2} \label{C_Sch},
\end{equation}
where $n$ and $m$ are the degeneracies of the ground and excited multiplets, respectively, and $\Delta$ an energy gap between the two multiplets. 
The calculations with the two-level model are shown with the dashed lines in Fig. \ref{fig_HC_Pr}(a), labeled ``2-3" (red), ``1-3" (blue), ``3-2" (green), and ``3-1" (yellow) for the doublet-triplet, singlet-triplet, triplet-doublet, and triplet-singlet models, respectively, with an energy gap of $\Delta$ $=$14 K. 
The (red) dashed curve of the doublet-triplet model better reproduces the $C_{\rm mag}(T)$ data than the calculations with the other three models. 
We confirmed that the doublet-triplet model gives still better fit to the $C_{\rm mag}$ data by taking the higher CEF levels into consideration.
In addition, taking account of the van-Vleck paramagnetic behavior of $\chi(T)$, the CEF ground state of the Pr ions is most likely to be the non-Kramers $\Gamma_{3}$ doublet for the $T_d$ point group. 
The magnetic entropy $S_{\rm mag}(T)$ estimated by integrating the $C_{\rm mag}$$/$$T$ data with respect to temperature is plotted with respect to the right axis of Fig. \ref{fig_HC_Pr}(a).
At around 3 K, $S_{\rm mag}(T)$ reaches $R$ln2, which supports the $\Gamma_{3}$ doublet ground state of the Pr ion.

Figure \ref{fig_HC_Pr}(b) shows the 4$f$ contribution to the specific heat, $C_{4f}$, in various constant magnetic fields of $B$ $\le$ 10 T.
We evaluated $C_{4f}$ by subtracting not only the phonon contribution of $C_{\rm ph}$ but also the nuclear contribution of a $^{141}$Pr nuclear spin with $I$ $=$ 5$/$2 from the measured $C(T)$ data.
The nuclear contribution was estimated by taking account of hyper-fine interaction between the $^{141}$Pr nuclear spin and the 4$f$ electrons \cite{Kondo61,Yamane18_AIP} and the Zeeman effect of magnetic fields on the Pr nuclear spin.
Looking at the data for $T$ $<$ 1 K, we notice a broad maximum at around 0.7 K.
The maximum is much smaller than the calculation with a singlet-singlet two-level model shown with the dotted (purple) line.
With increasing $B$, the maximum becomes broad and shifts to higher temperatures, and then reaches 1.3 K at 6 T. 
The weak field dependence of the maximum suggests that the maximum originates from the release of the magnetic entropy in the nonmagnetic $\Gamma_{3}$ doublet, e.g., short-range order of the quadrupolar degrees of freedom in the doublet.
The lack of the long-range order is probably attributed to small degrees of atomic disorder which could lift the degeneracy of the nonmagnetic doublet. 
The presence of intrinsic atomic disorder reflects in the small value of RRR $=$ 3.9 obtained from the $\rho(T)$ data shown in Fig. \ref{f_rho_01}.

As shown in the inset of Fig. \ref{fig_HC_Pr}(a), $C/T$ divergently increases on cooling below 0.2 K, where $\rho(T)$ increases on cooling (see the lower inset of Fig. 5).
In the anomalous behaviors of $C/T$ and $\rho(T)$, the $c$-$f$ hybridization may play a role to form the correlated electronic ground state with the multipolar degrees of freedom in the non-Kramers doublet ground state. 
This will be the subject to be addressed by further study.

\begin{figure}
\centering
\includegraphics[scale=0.34]{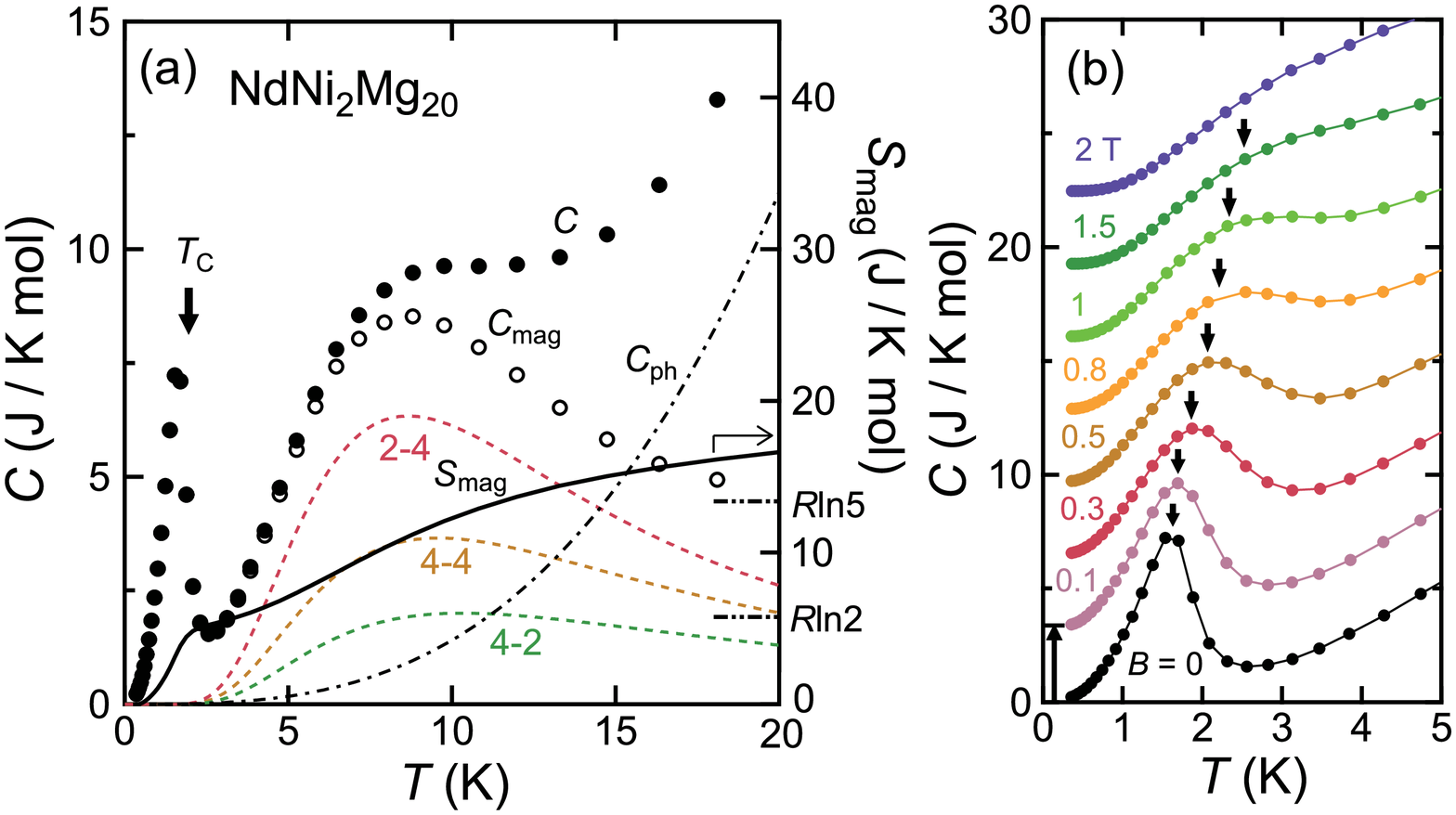}
\caption{(Color online) 
(a) Temperature dependence of the specific heat $C(T)$ and the magnetic entropy $S_{\rm mag}(T)$ of NdNi$_2$Mg$_{20}$.
The opened circles are the magnetic specific heat $C_{\rm mag}$. The dot-dashed line indicates the phonon contribution $C_{\rm ph}$ evaluated by the Debye model.
For comparison with the maximum of $C_{\rm mag}$ at 9 K, the dashed line labeled ``2-4" (red), ``4-4" (yellow), and ``4-2" (green) display calculations with two-level models of doublet-quartet, quartet-quartet, and quartet-doublet, respectively, with an energy gap of 23 K. 
A lambda-type anomaly peaking at 1.5 K is ascribed to the ferromagnetic transition.
Note that the arrow indicates $T_{\rm C}$ $=$ 1.9 K determined by the sharp bend in $\rho(T)$.
(b) $C(T)$ in magnetic fields of $B$ $\le$ 2 T.
With increasing $B$, the peak at 1.5 K at $B$ $=$ 0 becomes broad and shifts to higher temperatures, and then reaches 2.5 K at 1.5 T. 
The data are vertically offset for clarity.
}
\label{fig_HC_Nd}
\end{figure}

Figure \ref{fig_HC_Nd}(a) shows the $C(T)$ data of NdNi$_2$Mg$_{20}$. 
The open circles display the magnetic contribution to the specific heat, $C_{\rm mag}$, obtained by subtracting the phonon contribution $C_{\rm ph}$ from the $C(T)$ data.
$C_{\rm ph}$ is evaluated by the Debye model with $\theta_{\rm D}$ $=$ 317 K as described above.
We notice a broad maximum of $C_{\rm mag} (T)$ at $\sim$9 K and a sharp peak at 1.5 K.
The cubic CEF splits the ground $J$ $=$ 9$/$2 multiplet of the Nd ion into $\Gamma_6$ doublet and two $\Gamma_8$ quartets \cite{Lea62}.
Since the maximum at 9 K is better reproduced by the doublet-quartet two-level model with an energy gap of 23 K as shown with the (red) dashed curve than the other two models of quartet-quartet (yellow) and quartet-doublet (green),
we identify the CEF ground state as the Kramers $\Gamma_{6}$ doublet.
As similar to the case of PrNi$_2$Mg$_{20}$, the $C_{\rm mag}$ data are still better represented by taking account of the excitations to the higher CEF level.

The lambda-type anomaly peaking at 1.5 K indicates a magnetic phase transition.
It is noted that this peak temperature is lower than the transition temperature of $T_{\rm C}$ $=$ 1.9 K determined from the sharp bend of $\rho(T)$ shown in the upper inset of Fig. \ref{f_rho_01}.
This discrepancy probably results from short-range correlations between the magnetic moments of the Nd ions.
In fact, the $C(T)$ data exhibit a broad tail above the peak temperature.
As shown in Fig. \ref{fig_HC_Nd}(b), with increasing magnetic fields, the peak at 1.5 K becomes broad and shifts to higher temperatures, and reaches 2.5 K at 1.5 T.
This is a defining characteristic of the ferromagnetic transition, which becomes a crossover under applied magnetic fields.
The magnetic entropy $S_{\rm mag}(T)$ was estimated by integrating the $C_{\rm mag}/T$ data with respect to temperatures.
The value is 77\% of $R$ln2 at $T_{\rm C}$ $=$ 1.9 K and reaches $R$ln2 at 3 K.
Therefore, the ferromagnetic transition arises from the intersite magnetic interaction between the magnetic moments of the $\Gamma_{6}$ doublet ground state of the Nd ions.
The reduced magnetic entropy at $T_{\rm C}$ is probably ascribed to magnetic fluctuations above $T_{\rm C}$ due to the small degrees of atomic disorder.

\section{Summary}

In summary, we have prepared polycrystalline samples of $R$Ni$_2$Mg$_{20}$ ($R$ = Pr and Nd) and studied the structural, transport and magnetic properties. 
The powder XRD patterns of the samples annealed between 350 and 450$^{\circ}$C for 7 days can be indexed with the cubic CeCr$_{2}$Al$_{20}$-type structure.
The lattice parameters were refined to be $a$ $=$ 15.944(1) and 15.943(2) {\AA} for $R$ $=$ Pr and Nd, respectively, which are larger than those of the $R$$T_2$$X_{20}$ compounds with $X$ $=$ Al, Zn, and Cd.

The $\rho(T)$ data monotonically decrease on cooling from 300 to 40 K and exhibit shoulders at around 13 and 15 K for $R$ $=$ Pr and Nd, respectively. 
The trivalent states of the Pr and Nd ions are indicated by the effective magnetic moments of the Curie--Weiss behavior of $\chi(T)$ between 50 and 300 K.
The derived values for $\theta_{\rm p}$ are  $-$2.1 K and +2.9 K, respectively, for $R$ $=$ Pr and Nd, implying the antiferro- and ferro-type magnetic interactions. 

For $R$ $=$ Pr, $\chi(T)$ exhibits the van-Vleck-like paramagnetic behavior below 5 K.
A broad maximum in $C_{\rm mag}(T)$ at around 7 K is reproduced by a doublet-triplet two-level model with an energy gap of 14 K, suggesting the CEF ground state of the Pr ion is the non-Kramers $\Gamma_3$ doublet with quadrupolar degrees of freedom.
Because broad peak in $C_{\rm mag}$ at around 0.7 K cannot be reproduced by a singlet-singlet two-level model, we propose that the entropy of the $\Gamma_3$ doublet is released by short-range correlation between the quadrupole moments.
Small degrees of atomic disorder may lift the degeneracy of the nonmagnetic doublet to hinder the long-range order.

On the other hand, for $R$ $=$ Nd, $\chi(T)$ increases on cooling down to 1.8 K.
A broad maximum in $C_{\rm mag}(T)$ at around 9 K is reproduced by a doublet-quartet two-level model, indicating the Kramers $\Gamma_6$ doublet ground state with magnetic dipole moments for the Nd ion.
A sharp bend of $\rho(T)$ at 1.9 K and a lambda-type peak in $C(T)$ at 1.5 K suggest a phase transition among magnetic moments of the Nd ions in the Kramers doublet ground state.
With increasing magnetic fields, the peak in $C(T)$ becomes broad and shifts to higher temperatures, and then reaches 2.5 K at 1.5 T, which is a characteristic of the ferromagnetic transition.

\section*{Acknowledgments}

The authors would like to thank Y. Shimura, R. Yamamoto, K. T. Matsumoto, and K. Wakiya for helpful discussion.
The authors also thank Y. Shibata for the electron-probe microanalysis measurements carried out at N-BARD, Hiroshima University. 
The measurements with MPMS, PPMS and the mFridge were performed at N-BARD, Hiroshima University. 
We acknowledge support from Center for Emergent Condensed-Matter Physics (ECMP), Hiroshima University. 
This work was financially supported by grants in aid from MEXT/JSPS of Japan, Grants Nos. JP26707017, JP15H05886 (J-Physics), and JP18H01182.




\end{document}